\documentclass[aps,pra,eqsecnum,twocolumn,showpacs,superscriptaddress]{revtex4}
\usepackage[]{graphicx}
\usepackage{amsmath,multirow}
\begin{document}
\title{ Increasing and decreasing entanglement characteristics for continuous variables by a local photon subtraction}

\author{Su-Yong Lee}
\affiliation{Centre for Quantum Technologies, National University of Singapore, 3 Science Drive 2, 117543 Singapore, Singapore}
\affiliation{Department of Physics, Texas A\&M University at Qatar,
 Education City, POBox 23874, Doha, Qatar}

\author{Se-Wan Ji}
\affiliation{Department of Physics, Texas A\&M University at Qatar,
 Education City, POBox 23874, Doha, Qatar}
\affiliation{School of Computational Sciences, Korea Institute for Advanced Study, 207-43
Cheongryangri-dong, Dongdaemun-gu, Seoul 130-012, South Korea}

\author{Chang-Woo Lee}
\affiliation{Department of Physics, Texas A\&M University at Qatar,
 Education City, POBox 23874, Doha, Qatar}

\date{\today}

\begin{abstract}
We investigate how the entanglement characteristics of a non-Gaussian entangled state are increased or decreased by a local photon subtraction operation.
The non-Gaussian entangled state is generated by injecting a single-mode non-Gaussian state and a vacuum state into a 50:50 beam splitter.
We consider a photon-added coherent state $\hat{a}^{\dag}|\alpha\rangle$ and an odd coherent state $|\alpha\rangle-|-\alpha\rangle$ as
a single-mode non-Gaussian state.
In the regime of small $|\alpha|$, 
we show that the performance of quantum teleportation and the second-order Einstein-Podolsky-Rosen-type correlation can both be enhanced,
whereas the degree of entanglement decreases, for the output state when a local photon subtraction operation is applied to
the non-Gaussian entangled state.
The counterintuitive effect is more prominent in the limit of $|\alpha|\sim 0$.
\end{abstract}

\maketitle

\section{Introduction}
Entangled resources are useful in quantum teleportation, cryptography, and computation. 
In continuous variable (CV) regime, two-mode Gaussian entangled states are typically employed as entangled resources.
For two-mode Gaussian entangled states, entanglement can be fully described by  Einstein-Podolsky-Rosen (EPR) correlation \cite{Einstein}
which is characterized up to the second-order moments of the state \cite{Duan, Simon, Giedke}.
For two-mode non-Gaussian entangled states, however, entanglement is fully described with all orders of moments \cite{Shchukin,Biswas,Zubairy}.
Some of us have recently proposed entanglement criteria beyond the Gaussian regime, where the entanglement criteria including all orders of
EPR correlations can be measured with homodyne detection \cite{Nha}. 
Non-Gaussian entangled states provide the benefits on enhancing violation of Bell's inequality \cite{Nha1,Cerf,PLLN12},
and degree of entanglement \cite{Welsch,Cochrane,Olivares,Kitagawa,Dell,Yang1,Takahashi,Li,Suyong,Carlos}.

A certain class of non-Gaussian entangled state is simply generated by applying a non-Gaussian operation on a two-mode Gaussian state.
Typical non-Gaussian operations are photon addition and subtraction operations. 
The photon addition operation which was proposed \cite{Agarwal} was implemented \cite{Zavatta} 
 via a nondegenerate parametric amplifier with small coupling strength.
The photon subtraction operation was implemented \cite{Wenger} with a beam splitter of high transmittivity,
and considered in enhancing not only entanglement but also
performance of quantum-noise-limited amplifier \cite{Filip, Usuga, Jeffers, Hojoon}.
The sequential operations, such as photon addition-then-subtraction and subtraction-then-addition operations, 
were also studied \cite{Kim,Yang,Lee} and implemented in \cite{Parigi}.
In particular, the photon-addition-then-subtraction operation was considered in achieving a noiseless amplifier \cite{ZFB11},
quantifying bosonic behavior in composite particle system \cite{Pawel}, and distinguishing quantum particles from classical particles \cite{LNK13}. 
Based on an interferometric setting, the coherent superpositions of second-order operations, $\hat{a}\hat{a}^{\dag}\pm \hat{a}^{\dag}\hat{a}$,
was proposed \cite{Kim2} and implemented \cite{Zavatta1}. Moreover, it was also proposed in a cavity system \cite{Park}.
Some of us have recently proposed the coherent superposition of the elementary operation, $t\hat{a}+r\hat{a}^{\dag}$ \cite{SuYong},  as well as
other coherent superpositions of second-order operations, $t\hat{a}^2+r\hat{a}^{\dag 2}$ \cite{Yong} and 
$t\hat{a}\hat{a}^{\dag}+r\hat{a}^{\dag 2}$ \cite{Changwoo}.
Other coherent superposition operations, such as $t\hat{a}+r\hat{b}^{\dag}$ and $t\hat{a}\hat{a}^{\dag}+r\hat{a}^{\dag}\hat{a}$, were also proposed
to produce an arbitrary photon-number entangled state in a finite dimension, $\sum^{N}_{n=0}c_n|n,n\rangle_{AB}$ \cite{SJHH}.

Due to the fact that entanglement characteristics of Gaussian states are enhanced by a non-Gaussian operation 
\cite{Nha1,Cerf,PLLN12,Welsch,Cochrane,Olivares,Kitagawa,Dell,Yang1,Takahashi,Li,Suyong,Carlos}, 
it is natural to have a question about whether entanglement characteristics of non-Gaussian states are enhanced by a non-Gaussian operation.
In particular, we are interested in non-Gaussian states which do not have any two-mode squeezing properties in order to determine their own usefulness compared with a typical Gaussian entangled state.
Then, we apply a simple non-Gaussian operation, i.e., local photon subtraction operation, to the non-Gaussian states.
We generate the non-Gaussian entangled state by injecting a single-mode non-Gaussian state and a vacuum state into a 50:50 beam splitter,
where we consider a photon-added coherent state $\hat{a}^{\dag}|\alpha\rangle$  and an odd coherent state $|\alpha\rangle-|-\alpha\rangle$ as the single-mode
non-Gaussian state. 
With these non-Gaussian entangled states, we investigate the entanglement characteristics: Degree of entanglement,
second-order Einstein-Podolsky-Rosen (EPR) correlation, and performance of quantum teleportation in  Braunstein and Kimble (BK)'s protocol \cite{Braunstein}.
After a local photon subtraction operation on the non-Gaussian resources in the regime of small $|\alpha |$,
we find that the teleportation fidelity of a coherent state and the second-order EPR correlation are enhanced whereas
the degree of entanglement diminishes. 
In the limit of $|\alpha|\sim 0$, the counterintuitive effect is more prominent: The teleportation fidelity increases from $0.25$ to beyond the classical limit $0.5$, 
and the second-order EPR correlation begins to emerge, whereas the degree of entanglement decreases from $1$ to $0$.
 
In this paper, we begin in Sec. II with the generation of a non-Gaussian entangled state with a 50:50 beam splitter, injecting
a single-mode non-Gaussian state and a vacuum state. In Sec. III we investigate the entanglement properties (entanglement and second-order EPR correlation) 
of the non-Gaussian entangled state via local photon subtraction operation. Then, we employ the non-Gaussian entangled state 
via local photon subtraction operation for CV teleportation in Sec. IV.
The main results are summarized in Sec. V.

\section{Generation of non-Gaussian entangled states}
We generate a non-Gaussian entangled state by injecting a single-mode non-Gaussian state $|\psi\rangle$ 
and a vacuum state $|0\rangle$ into a 50:50 beam splitter. 
We consider a photon-added coherent state
$\hat{a}^{\dag}|\alpha\rangle$, and an odd coherent state $|\alpha\rangle-|-\alpha\rangle$ as the single-mode non-Gaussian state.
We can simply check that the single-mode non-Gaussian states are maximally nonclassical due to the fact :
Given $|\langle \beta |\psi\rangle|^2=0$ for at least one coherent state $|\beta\rangle$, then the state is maximally nonclassical  \cite{Nobert}.
The photon-added coherent state was implemented in the laboratory \cite{Zavatta}. To generate the odd coherent state was proposed in some ways \cite{Yurke,Mecozzi,Marek,Changwoo}, and implemented in other ways \cite{Polzik, Grangier1, Grangier2, Sasaki1, Sasaki2, Grangier3, Gerrits}.
Applying local photon subtraction operations on the non-Gaussian entangled state, 
we can simply obtain the following form,
\begin{eqnarray}
|\Psi\rangle_{AB}&=&\frac{1}{\sqrt{N}}\hat{a}^n\hat{b}^m\hat{B}_{AB}|\psi\rangle_A|0\rangle_B \nonumber\\
&=&\frac{1}{\sqrt{N}} \hat{B}_{AB}(\frac{1}{\sqrt{2}})^{n+m}\hat{a}^{n+m}|\psi\rangle_A|0\rangle_B,
\end{eqnarray}
where $N$ is a normalization factor, and the beam splitting transformation is $\hat{B}^{\dag}_{AB}\hat{a}\hat{B}_{AB}= \frac{1}{\sqrt{2}}(\hat{a}-\hat{b}), ~
\hat{B}^{\dag}_{AB}\hat{b}\hat{B}_{AB}= \frac{1}{\sqrt{2}}(\hat{b}+\hat{a})$. The values of $n$ and $m$ are non-negative integers.

In the case of $|\psi\rangle_A=\hat{a}^{\dag}|\alpha\rangle$, the output state $|\Psi\rangle_{AB}$ becomes 
\begin{eqnarray}
|\Psi_1\rangle_{AB}=\frac{1}{\sqrt{N_1}}[n+m+\frac{\alpha}{\sqrt{2}}(\hat{a}^{\dag}-\hat{b}^{\dag})]
|\frac{\alpha}{\sqrt{2}},\frac{-\alpha}{\sqrt{2}}\rangle_{AB},\nonumber\\
\end{eqnarray}
where $N_1=(n+m+|\alpha|^2)^2+|\alpha|^2$.
In the case of $|\psi\rangle_A=|\alpha\rangle-|-\alpha\rangle$, the output state $|\Psi\rangle_{AB}$ becomes
\begin{eqnarray}
|\Psi_2\rangle_{AB}=\frac{1}{\sqrt{N_2}}[|\frac{\alpha}{\sqrt{2}},\frac{-\alpha}{\sqrt{2}}\rangle_{AB}-(-1)^{n+m}|\frac{-\alpha}{\sqrt{2}},\frac{\alpha}{\sqrt{2}}\rangle_{AB}],\nonumber\\
\end{eqnarray}
where $N_2=2[1-(-1)^{n+m}e^{-2|\alpha|^2}]$. According to the total number of the photon subtraction operation, the relative phase becomes plus (minus) at odd (even) number 
because the state $|\psi\rangle_A=|\alpha\rangle-|-\alpha\rangle$ is transformed into an even (odd) coherent state by local photon subtraction operations, as shown in Eq. (2.1).
In the next section, we investigate entanglement properties of these non-Gaussian entangled states after they are processed with local photon subtraction operation. 

\section{Entanglement properties}

\subsection{Entanglement}
Entanglement for a pure bipartite state is described with 
von Neumann entropy calculated as 
$E=-Tr[\rho_{A}\log_2{\rho_{A}}]=-Tr[\rho_{B}\log_2{\rho_{B}}]
=-\sum_i \lambda_i\log_2{\lambda_i}$, ($\lambda_i$: eigenvalues of $\rho_{A}$
or $\rho_{B}$) \cite{BBPS}.
For $|\psi\rangle_A=\hat{a}^{\dag}|\alpha\rangle$, the entangled state $|\Psi_1\rangle_{AB}$ can be written as
\begin{eqnarray}
&&|\Psi_1\rangle_{AB}\nonumber\\
&&=\frac{1}{\sqrt{N_1}}[M_1|\frac{\alpha}{\sqrt{2}},0\rangle_A\otimes|\frac{-\alpha}{\sqrt{2}},0\rangle_B\nonumber\\
&&+\frac{\alpha}{\sqrt{2}}(|\frac{\alpha}{\sqrt{2}},1\rangle_A\otimes|\frac{-\alpha}{\sqrt{2}},0\rangle_B
-|\frac{\alpha}{\sqrt{2}},0\rangle_A\otimes|\frac{-\alpha}{\sqrt{2}},1\rangle_B)],\nonumber\\
\end{eqnarray}
where  $M_1=n+m+|\alpha|^2$, and we considered the relation, 
$\hat{D}^{\dag}(\alpha)\hat{a}^{\dag}\hat{D}(\alpha)=\hat{a}^{\dag}+\alpha^*$. 
We denote that
$|\frac{\alpha}{\sqrt{2}},0\rangle\equiv\hat{D}(\frac{\alpha}{\sqrt{2}})|0\rangle$, and
$|\frac{\alpha}{\sqrt{2}},1\rangle\equiv\hat{D}(\frac{\alpha}{\sqrt{2}})|1\rangle$ are displaced Fock states.
The displaced Fock states are orthonormal to each other,
such that we can get the following reduced density matrix,
\begin{eqnarray}
\rho_{|\Psi_1\rangle}=\frac{1}{2N_1}
\begin{pmatrix} |\alpha|^2+2M^2_1 & \sqrt{2}\alpha^* M_1 \\
\sqrt{2}\alpha M_1 & |\alpha|^2 \end{pmatrix}.
\end{eqnarray}
With the eigenvalues of the Eq. (3.2), we derive the degree of entanglement as a function of $|\alpha|$ in Fig. 1 (a).
The degree of entanglement decreases with the total 
number of the local photon subtraction operation ($n+m$).  In the limit of $|\alpha| \sim 0$, the degree of entanglement is changed 
from $1$ to nearly zero via a local photon subtraction operation.
\begin{figure}
\centerline{\scalebox{0.3}{\includegraphics[angle=270]{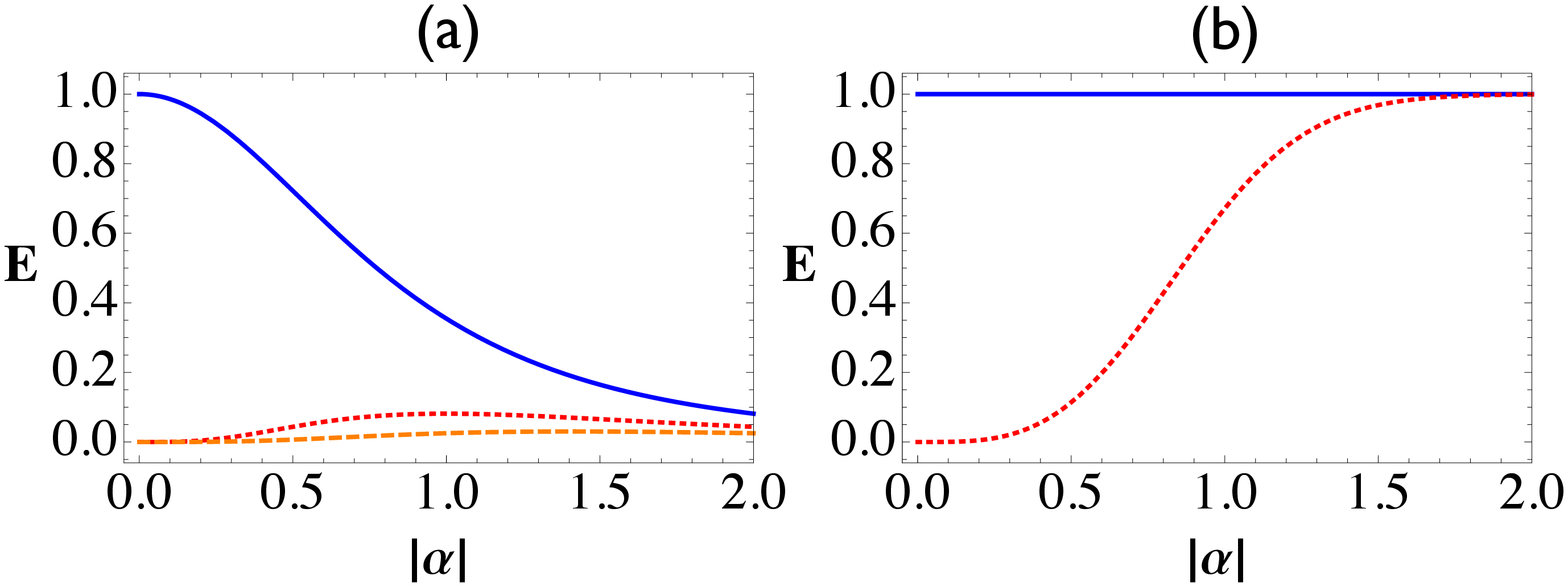}}}
\vspace{-1.2in}
\caption{Degree of entanglement for the non-Gaussian entangled state Eq. (2.1) as a function of $|\alpha|$. 
(a) $|\psi\rangle_A=\hat{a}^\dag|\alpha\rangle$:
$n+m=$ $0$ (blue solid), $1$ (red dotted), and $2$ (orange dashed),
(b) $|\psi\rangle_A=|\alpha\rangle-|-\alpha\rangle$:
$n+m=$ even (blue solid), and odd (red dotted)}
\label{fig:fig1}
\end{figure}

For $|\psi\rangle_A=|\alpha\rangle-|-\alpha\rangle$, 
the entangled state $|\Psi_2\rangle_{AB}$ can be written as
\begin{eqnarray}
&&|\Psi_2\rangle_{AB}\nonumber\\
&&=\frac{1}{2\sqrt{N_2}}[(1+e^{-|\alpha|^2})(1-(-1)^{n+m})|e,e\rangle_{AB}\nonumber\\
&&-(1-e^{-|\alpha|^2})(1-(-1)^{n+m})|o,o\rangle_{AB}\nonumber\\
&&+\sqrt{1-e^{-2|\alpha|^2}}(1+(-1)^{n+m})(|o,e\rangle_{AB}-|e,o\rangle_{AB})],\nonumber\\
\end{eqnarray}
where $|o\rangle\equiv\frac{1}{\sqrt{2(1-e^{-|\alpha|^2})}}
(|\frac{\alpha}{\sqrt{2}}\rangle-|\frac{-\alpha}{\sqrt{2}}\rangle)$ is an odd coherent state, and 
$|e\rangle\equiv\frac{1}{\sqrt{2(1+e^{-|\alpha|^2})}}
(|\frac{\alpha}{\sqrt{2}}\rangle+|\frac{-\alpha}{\sqrt{2}}\rangle)$ is an even coherent state.
The odd and even coherent states are orthonormal to each other, such that 
the reduced density matrix is represented by
\begin{eqnarray}
\rho_{|\Psi_2\rangle}=\frac{1}{N_2}
\begin{pmatrix} \lambda_+  & 0\\ 
0 & \lambda_- \end{pmatrix},
\end{eqnarray}
where $\lambda_\pm=(1\pm e^{-|\alpha|^2})[1\mp(-1)^{n+m}e^{-|\alpha|^2}]$.
With the eigenvalues of Eq. (3.4), the degree of entanglement is derived as a function of $|\alpha|$ in Fig. 1 (b).
The degree of entanglement decreases from $1$ to less than or equal to $1$, via a local photon subtraction operation.
 In the case of the even number of local photon subtraction operations, the degree of entanglement
is constant as $1$. In the case of the odd number, it
 increases from $0$ to $1$ with $|\alpha|$. 
 
The above results may be understood by looking into the states in the regime of small $|\alpha|$.
For the state $\hat{a}^{\dag}|\alpha\rangle$, we consider it up to the first order of $\alpha$,
and then the non-Gaussian entangled state can be approximately written as  
\begin{eqnarray}
|\Psi_1\rangle_{AB}&\approx&|1,0\rangle_{AB}-|0,1\rangle_{AB}+\alpha(|2,0\rangle_{AB}\nonumber\\
&&+|0,2\rangle_{AB}-\sqrt{2}|1,1\rangle_{AB}), 
\end{eqnarray}
where $n+m=0$ and the state is not normalized. After a local photon subtraction operation on either of the modes,  Eq. (3.5) is transformed into
\begin{eqnarray}
\hat{a}|\Psi_1\rangle_{AB}\approx|0,0\rangle_{AB}+\sqrt{2}\alpha(|1,0\rangle_{AB}-|0,1\rangle_{AB}).
\end{eqnarray}
In the limit of $|\alpha| \rightarrow 0$, Eq.(3.5) goes to a maximally entangled state but
Eq.(3.6) goes to a separable state. 
For the state $|\alpha\rangle-|-\alpha\rangle$, we consider it up to the third order of $\alpha$, and then the non-Gaussian entangled state can be approximately written as
\begin{eqnarray}
|\Psi_2\rangle_{AB}&\approx&|1,0\rangle_{AB}-|0,1\rangle_{AB}+\frac{\alpha^2}{2\sqrt{6}}(|3,0\rangle_{AB}-|0,3\rangle_{AB}\nonumber\\
&&-\sqrt{3}|2,1\rangle_{AB}+\sqrt{3}|1,2\rangle_{AB}),
\end{eqnarray}
where $n+m=0$ and the state is not normalized. After a local photon subtraction operation on either of the modes,  Eq. (3.7) is transformed into
\begin{eqnarray}
\hat{a}|\Psi_2\rangle_{AB}&\approx&|0,0\rangle_{AB}+\frac{\alpha^2}{2\sqrt{2}}(|2,0\rangle_{AB}+|0,2\rangle_{AB}\nonumber\\
&&-\sqrt{2}|1,1\rangle_{AB}).
\end{eqnarray} 
In the limit of $|\alpha| \rightarrow 0$, Eq.(3.7) goes to a maximally entangled state but Eq. (3.8) goes to a separable state.

\subsection{Second-order Einstein-Podolsky-Rosen correlation}
Second-order EPR correlation is described with the total variance of a pair of EPR-like operators,
\begin{eqnarray}
&&\Delta^2(\hat{x}_A-\hat{x}_B)+\Delta^2(\hat{p}_A+\hat{p}_B)\nonumber\\
&&=1+(\langle \hat{a}^{\dag}\hat{a}\rangle+\langle \hat{b}^{\dag}\hat{b}\rangle-\langle \hat{a}\hat{b}\rangle
-\langle \hat{a}^{\dag}\hat{b}^{\dag}\rangle)\nonumber\\
&&-(\langle\hat{a}\rangle-\langle\hat{b}^{\dag}\rangle)(\langle\hat{a}^{\dag}\rangle-\langle\hat{b}\rangle),
\end{eqnarray}
where $\hat{x}_j=\frac{1}{2}(\hat{a}_j+\hat{a}^{\dag}_j)$
and $\hat{p}_j=\frac{-i}{2}(\hat{a}_j-\hat{a}^{\dag}_j)$ $(j=A,B)$. 
The total variance which is less than $1$ indicates Gaussian quantum entanglement \cite{Duan}, an important resource in CV quantum protocols.
Given a symmetric state, we can evaluate the second-order EPR correlation with the expectation values such as
$\langle\hat{a}^{\dag}\hat{a}\rangle$, $\langle\hat{a}\hat{b}\rangle$, and $\langle\hat{a}\rangle$.
The other terms are obtained with the complex conjugate of the expectation value, e.g., $\langle\hat{a}^{\dag}\rangle=\langle\hat{a}\rangle^{*}$. 

In the case of $|\psi\rangle_A=\hat{a}^{\dag}|\alpha\rangle$, the  second-order EPR correlation of Eq. (2.2) is described in this form,
\begin{eqnarray}
&&\Delta^2(\hat{x}_A-\hat{x}_B)+\Delta^2(\hat{p}_A+\hat{p}_B)\nonumber\\
&&=1+\frac{|\alpha|^2[(M_1+1)^2+|\alpha|^2](1+\cos{2\varphi})}{N_1}\nonumber\\
&&-\frac{|\alpha|^2\cos{2\varphi}}{N_1}-\frac{2|\alpha|^2(N_1+M_1)^2\cos^2{\varphi}}{N^2_1},
\end{eqnarray}
where $\alpha=|\alpha|e^{i\varphi}$. Equation (3.10) is optimized at $\varphi=0$.
We consider the second-order EPR correlation as a function of $|\alpha|$ at $n+m=0,1,2$, as shown in Fig. 2 (a).  
At $n+m=0$, the second-order EPR correlation is not shown in the region of $|\alpha| \leq1$. 
However, applying local photon subtraction operations on the state, we can see that the EPR correlation shows up in the region of $|\alpha|\leq 1$ at $n+m=1,2$.
In the region of $|\alpha|\leq 1.454$, the second-order  EPR correlation is improved by a local photon subtraction operation.
In the case of $|\psi\rangle_A=|\alpha\rangle-|-\alpha\rangle$, the second-order EPR correlation of Eq. (2.3) is described in this form,
\begin{eqnarray}
&&\Delta^2(\hat{x}_A-\hat{x}_B)+\Delta^2(\hat{p}_A+\hat{p}_B)\nonumber\\
&&=1+|\alpha|^2[\frac{1+(-1)^{n+m}e^{-2|\alpha|^2}}{1-(-1)^{n+m}e^{-2|\alpha|^2}}+\cos{2\varphi}],
\end{eqnarray}
where $\alpha=|\alpha|e^{i\varphi}$. Equation (3.11) is optimized at $\varphi=\pi/2$.
We consider the second-order EPR correlation as a function of $|\alpha|$ at $n+m=even/odd$, as shown in Fig. 2(b).
At $n+m=even$, the second-order EPR correlation is not shown in the whole region of $|\alpha|$. 
On the other hand, the second-order EPR correlation shows up in the whole region of $|\alpha|$ at $n+m=odd$. 
\begin{figure}
\centerline{\scalebox{0.3}{\includegraphics[angle=270]{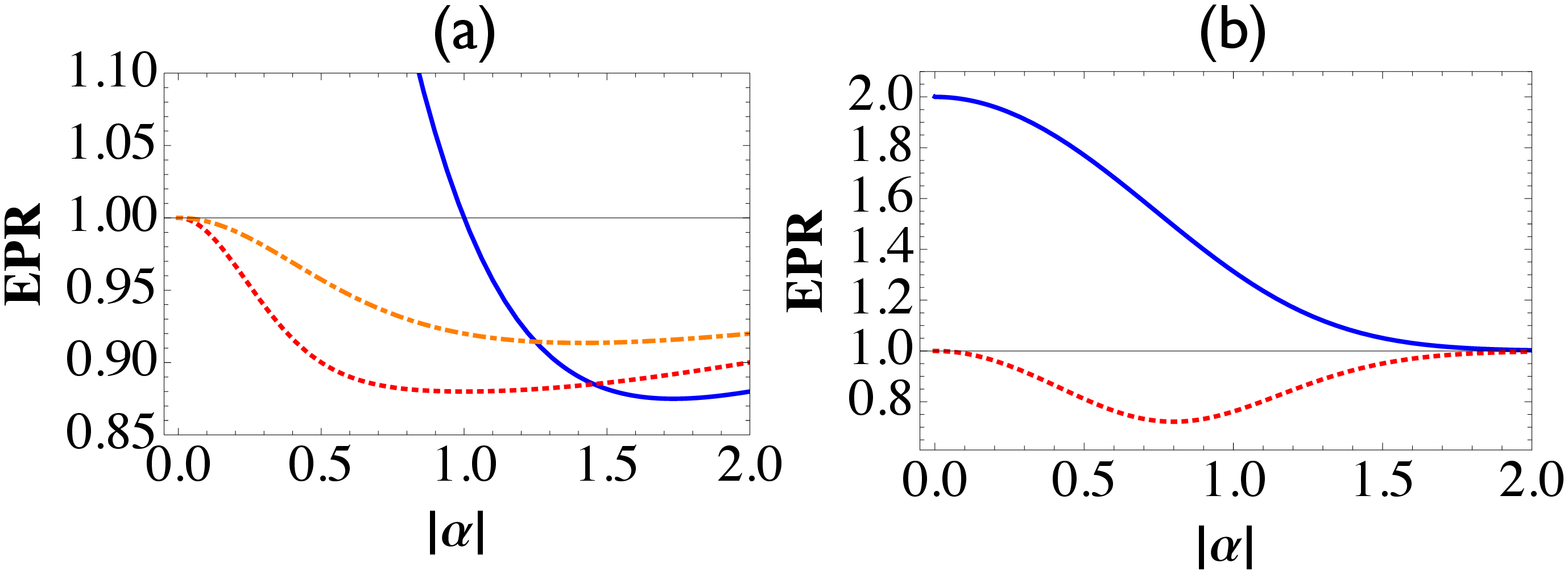}}}
\vspace{-1.2in}
\caption{ Second-order EPR correlation of the non-Gaussian entangled state Eq. (2.1) as a function of $|\alpha |$.
 (a) $|\psi\rangle_A=\hat{a}^\dag|\alpha\rangle$:
$n+m=$ $0$ (blue solid), $1$ (red dotted), and $2$ (orange dashed),
(b) $|\psi\rangle_A=|\alpha\rangle-|-\alpha\rangle$ :
$n+m=$ even (blue solid), and odd (red dotted)}
\label{fig:fig2}
\end{figure}

From the above results, we could understand that local photon subtraction operation plays the role of second-order EPR correlation for the non-Gaussian entangled states. 

\section{Continuous variable teleportation using non-Gaussian entangled states}
After a local photon subtraction operation on the non-Gaussian entangled states, 
we could see that the second-order EPR correlation can be created whereas the degree of entanglement decreases.
Now we consider the non-Gaussian entangled states as entangled resources to teleport a coherent state in continuous variable (CV) teleportation.
We consider Braunstein and Kimble (BK)'s protocol \cite{Braunstein} whose performance is evaluated by the average fidelity 
between an unknown input state and its teleported state.
Teleportation of coherent states has a classical limit of average fidelity
$F_{classical}=1/2$ if Alice and Bob make use of a classical channel \cite{Braunstein2}.
The average fidelity of teleportation is given by
\begin{eqnarray}
F=\frac{1}{\pi}\int d^2\lambda C_{out}(\lambda)C_{in}(-\lambda),
\end{eqnarray}
where $C_{out}(\lambda)=C_{in}(\lambda)C_{E}(\lambda^*,\lambda)$ \cite{Marian} is the characteristic function of the teleported state. 
Here, $C_{E}(\lambda^*,\lambda)$ is the characteristic function of an entangled resource, 
and $C_{in}(\lambda)$ is one of an input state. 
For input coherent states, it is sufficient to calculate the teleportation fidelity for a particular input coherent state \cite{Suyong}
since there is no difference between the amplitudes of the input and output coherent states in BK protocol.
For brevity, we will refer to fidelity as the average fidelity hereafter.

In the case of $|\psi\rangle_A=\hat{a}^{\dag}|\alpha\rangle$,  we consider the non-Gaussian entangled state Eq.(2.2) to teleport a coherent state.
The characteristic function of the state is given by
\begin{eqnarray}
C_E(\lambda_2,\lambda_3)&=&\frac{e^{-\frac{1}{2}(|\lambda_2|^2+|\lambda_3|^2)+\delta-\delta^*}}{N_1}[~|\alpha|^2 \nonumber\\
&&+(M_1+\delta)(M_1-\delta^*)~],
\end{eqnarray}
where $\delta=\frac{\alpha^*}{\sqrt{2}}(\lambda_2-\lambda_3)$.
Using Eq. (4.1), we can obtain the teleportation fidelity, which is optimized at $\varphi=0$ for $\alpha=|\alpha |e^{i\varphi}$.
We consider the teleportation fidelity as a function of $|\alpha|$ at $n+m=0,1,2$, as shown in Fig. 3 (a). 
At $n+m=0$, the teleportation fidelity is less than $1/2$ in the region of $|\alpha| < 0.686$.
At $n+m=1,2$, the teleportation fidelity becomes larger than $1/2$ in the whole region of $|\alpha|$.
The teleportation fidelity is improved by a local photon subtraction operation in the region of $|\alpha|<0.963$.
In the case of $|\psi\rangle_A=|\alpha\rangle-|-\alpha\rangle$, we obtain the following characteristic function,
\begin{eqnarray}
C_E(\lambda_2,\lambda_3)&=&\frac{2e^{-(|\lambda_2|^2+|\lambda_3|^2)/2}}{N_2}[~\cosh(\delta-\delta^*)\nonumber\\
&&-(-1)^{n+m}e^{-2|\alpha|^2}\cosh(\delta+\delta^*)]. \nonumber\\
\end{eqnarray}
Using Eq. (4.1), we can obtain the teleportation fidelity which is optimized at $\varphi=\pi/2$ for $\alpha=|\alpha|e^{i\varphi}$.
We consider the teleportation fidelity as a function of $|\alpha|$ at $n+m= even/odd$, as shown in Fig. 3 (b). 
The teleportation fidelity becomes larger than $1/2$ in the whole region of $|\alpha|$ by an odd number of local photon subtraction operation.

We may understand the above result by comparing with the second-order EPR correlation and the degree of entanglement. 
First of all, we compare the teleportation fidelity with the second-order EPR correlation in Figs. 2 and 3. 
In the case of $|\psi\rangle_A=\hat{a}^{\dag}|\alpha\rangle$, the teleportation fidelity, which is larger than $1/2$, does
not guarantee the existence of the second-order EPR correlation in the region of $0.686\leq |\alpha|<1$.
In the case of $|\psi\rangle_A=|\alpha\rangle-|-\alpha\rangle$, on the other hand, 
the teleportation fidelity which is larger than $1/2$  guarantees the existence of the second-order EPR correlation.
The former case can be explained by all orders of EPR correlation, such that
we consider another teleportation fidelity formula represented by all orders of EPR correlation, 
$F_{epr}=\langle e^{-\Delta^2\hat{u}-\Delta^2\hat{v}}\rangle_{\rho_{AB}}$ \cite{Nha}, 
where $\Delta^2\hat{u}+\Delta^2\hat{v}=[\hat{b}^{\dag}-\langle\hat{b}^{\dag}\rangle-\hat{a} +\langle\hat{a}\rangle] 
[\hat{b}-\langle\hat{b}\rangle-\hat{a}^{\dag} +\langle\hat{a}^{\dag}\rangle]$.
Since we consider entangled states generated by a 50:50 beam splitter, the fidelity is simply transformed into \cite{Nha}
\begin{eqnarray}
F_{epr}=\langle e^{-2(\hat{X}_b-\langle \hat{X}_b\rangle)^2}\rangle\langle e^{-2(\hat{P}_a-\langle\hat{P}_a\rangle)^2} \rangle,
\end{eqnarray}
where $\hat{X}_b=\frac{1}{2}(\hat{b}+\hat{b}^{\dag})$ and $\hat{P}_a=\frac{-i}{2}(\hat{a}-\hat{a}^{\dag})$.
According to Eq. (2.1), we can derive the expectation value $\langle 0|e^{-2(\hat{X}_b-\langle \hat{X}_b\rangle)^2}|0\rangle=1/\sqrt{2}$
for the input mode B.
For the input mode A, we can employ the following relation,
\begin{eqnarray}
&&e^{-2(\hat{P}_a-\langle\hat{P}_a\rangle)^2}\nonumber\\
 &&=\frac{e^{2\langle\hat{P}_a\rangle^2}}{\sqrt{2}}e^{\frac{1}{4}\hat{a}^{\dag 2}}(\frac{1}{2})^{\hat{a}^{\dag}\hat{a}}
e^{\frac{1}{4}\hat{a}^{ 2}}\hat{D}(2i\langle \hat{P}_a\rangle) e^{-4i\langle \hat{P}_a\rangle \hat{a}}.
\end{eqnarray}
Thus, in the case of $|\psi\rangle_A=\hat{a}^{\dag}|\alpha\rangle$, we can get the following fidelity 
\begin{eqnarray}
F_{epr}=\frac{M(M+|\alpha\beta |)+\frac{|\alpha|^2}{2}(1+\frac{|\beta|^2}{2})}{2N_1}e^{-(|\alpha|-\frac{|\beta|}{2})^2},
\end{eqnarray} 
where $|\beta|=2|\alpha|(1+\frac{M+|\alpha|^2}{N_1})$ and $M=n+m$.
We can check that the teleportation fidelity $F_{epr}$ is equal to the fidelity obtained from Eq. (4.2). 
Therefore, the teleportation fidelity in BK protocol can be explained by all orders of the EPR correlation.
Second, we compare the teleportation fidelity with the degree of entanglement in Figs. 1 and 3. 
For both cases of $|\psi\rangle_A=\hat{a}^{\dag}|\alpha\rangle$ and $|\psi\rangle_A=|\alpha\rangle-|-\alpha\rangle$ in the regime of small $|\alpha|$, 
a local photon subtraction operation enhances the teleportation fidelity whereas the operator diminishes the degree of entanglement.
In the limit of $|\alpha|\sim 0$, the teleportation fidelity increases from $1/4$ to beyond $1/2$ 
whereas the degree of entanglement ($E$) decreases from $E=1$ to $E \sim 0$.
We can predict that all orders of the correlation in the non-Gaussian resources are not always useful to enhance the teleportation fidelity in BK protocol. 

\begin{figure}
\centerline{\scalebox{0.31}{\includegraphics[angle=270]{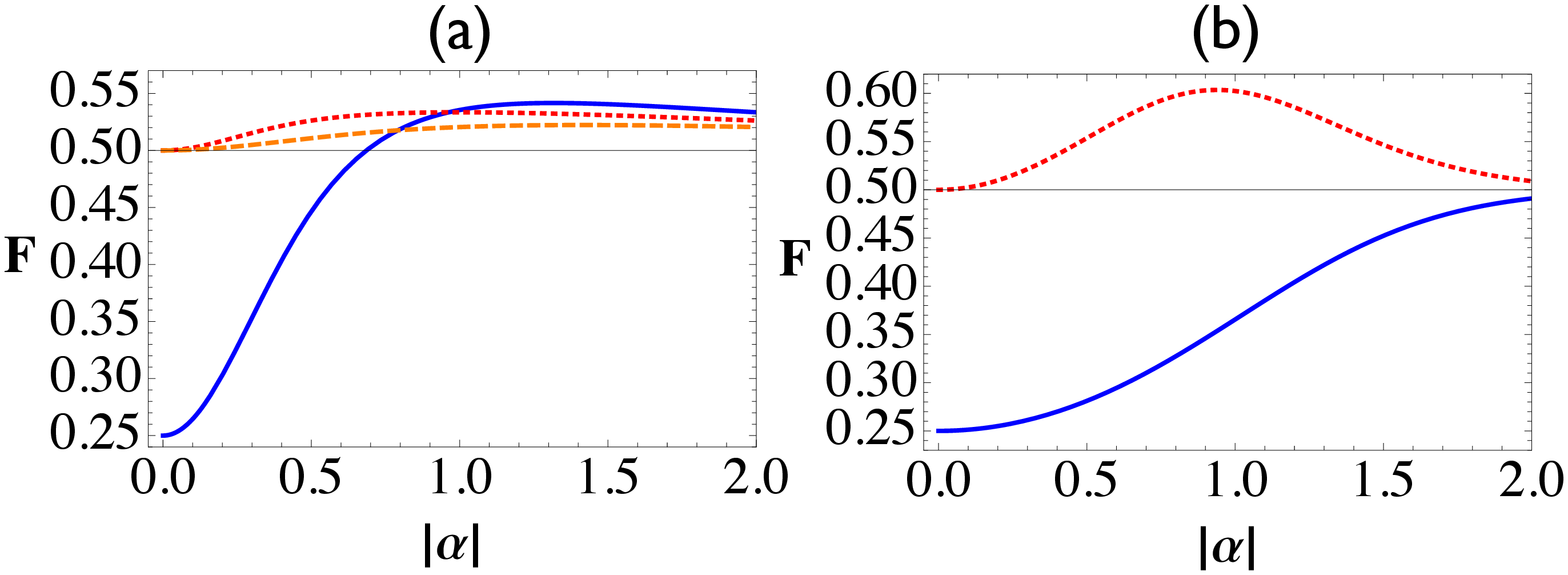}}}
\vspace{-1.3in}
\caption{ Teleportation fidelity of a coherent state  with the non-Gaussian entangled resource Eq. (2.1) as a function of $|\alpha|$.
(a) $|\psi\rangle_A=\hat{a}^\dag|\alpha\rangle$ : 
 $n+m=$ $0$ (blue solid), $1$ (red dotted), and $2$ (orange dashed),
(b) $|\psi\rangle_A=|\alpha\rangle-|-\alpha\rangle$:
 $n+m=$ even (blue solid), and odd (red dotted)}
\label{fig:fig3}
\end{figure}

Furthermore, we study some cases about the teleportation fidelity of classical limit $1/2$ regarding entanglement and second-order EPR correlation
of pure bipartite entangled states, as shown in table I.
Some entangled states with second-order EPR correlation can teleport a coherent state with the fidelity which is beyond $1/2$ \cite{Welsch, Olivares, Kitagawa}.
For the entangled states, $\hat{a}^{\dag}\hat{b}^{\dag}|TMSS\rangle$ and $\hat{a}^{\dag}\hat{b}^{\dag}\hat{a}\hat{b}|TMSS\rangle$, 
($|TMSS\rangle=\sqrt{1-\lambda^2}\sum^{\infty}_{n=0}\lambda^n|n\rangle_A|n\rangle_B$ ($\lambda=\tanh{s}$): two-mode squeezed vacuum state), 
in the region of $0.16<\lambda <0.4$, however,
 we can find that the second-order EPR correlation cannot be exhibited but the teleportation fidelity of a coherent state can be beyond $1/2$ \cite{Yang1}.
 There is another example shown in the supplemental material of the paper \cite{Nha}, where a single photon state can be teleported with the fidelity up to $1$
 via an entangled resource with no second-order EPR correlation.
For the entangled state, $\hat{a}|TMSS\rangle$, in the region of $\lambda\sim 0.38$, there is entanglement as well as second-order EPR correlation, 
but the teleportation fidelity of a coherent state can be below $1/2$ \cite{Suyong}.
For the entangled state at $n+m=0$ in Eq. (2.3), there is a high degree of entanglement without second-order EPR correlation, but the teleportation fidelity cannot 
be beyond $1/2$. For any pure bipartite separable states, the teleportation fidelity is below $1/2$.

\begin{table}[]
\caption{Teleportation fidelity in BK protocol} \label{tab: title}
\begin{tabular}{|c|c|c|c|}
\hline  Fidelity & Entanglement  & 2nd-order EPR  & case \\
\hline\hline  \multirow{2}{*} {$F>\frac{1}{2}$} & Yes & Yes & \cite{Welsch, Olivares, Kitagawa} \\
&Yes & No & \cite{Nha, Yang1},  this paper\\
\hline \multirow{3}{*} {$F<\frac{1}{2}$} & Yes & Yes & \cite{Suyong} \\
& Yes & No & this paper \\
& No & No & pure separable  \\
\hline
\end{tabular}
\end{table}

\section{Conclusion}
In this paper, we have shown that a local photon subtraction operation on a non-Gaussian entangled state, in the regime of small $|\alpha|$, can enhance 
the teleportation fidelity and the second-order EPR correlation while diminishing the degree of entanglement at the same time.
We considered the non-Gaussian entangled state generated by injecting a vacuum state and a photon-added coherent state $\hat{a}^{\dag}|\alpha\rangle$
 (an odd coherent state $|\alpha\rangle-|-\alpha\rangle$) into a 50:50 beam splitter.
In the limit of $|\alpha|\sim 0$, the local photon subtraction operation enhanced the teleportation fidelity by a little bit more than $1/2$ from $1/4$,
and made the second-order EPR correlation appear from nonexistence, 
whereas the degree of entanglement is reduced from $1$ to $0$ via the local photon subtraction operation.
In the regime of large $|\alpha|$, all the entanglement properties we considered slightly decreased via a local photon subtraction operation.
Furthermore, we could find the particular cases when the teleportation fidelity can be beyond (below) the classical limit $1/2$ without (with) second-order EPR correlation.

The present study can be compared with a Gaussian entangled state via a local photon subtraction operation.
We consider a two-mode squeezed vacuum state as the Gaussian entangled state.
We can find the opposite behavior of the Gaussian entangled state and the non-Gaussian entangled state we considered in this paper.
The photon subtraction operation on a two-mode squeezed vacuum state, $\hat{a}|TMSS\rangle$, enhances the degree of entanglement, 
but decreases the teleportation fidelity and the second-order EPR correlation \cite{Suyong}. 
In the small squeezing regime, the teleportation fidelity $F$ decreases from $F>1/2$ to $F<1/2$,
and the second-order EPR correlation goes from existence to nonexistence.

According to a quantum protocol, we can apply an appropriate local operation on an entangled state to enhance the performance. In our case, non-Gaussian states via a local photon subtraction operation
are useful for performing the BK protocol. As a further work, it would be interesting to investigate what kind of local operation on an entangled state is appropriate to enhance the performance of some quantum protocols. 

\begin{acknowledgments}
The authors would like to thank Prof. Hyunchul Nha for his scientific guidance given in discussions.
S.Y.L. is supported by the National Research Foundation and Ministry of Education in Singapore.
\end{acknowledgments}


\begin{thebibliography}{99}

\bibitem{Einstein} A. Einstein, B. Podolsky, and N. Rosen, Phys. Rev. \textbf{47}, 777 (1935).

\bibitem{Duan} L.M. Duan, G. Giedke, J.I. Cirac, and P. Zoller, \prl \textbf{84}, 2722 (2000).

\bibitem{Simon} R. Simon, \prl \textbf{84}, 2726 (2000).

\bibitem{Giedke} G. Giedke, B. Kraus, M. Lewenstein, and J.I. Cirac, \prl \textbf{87}, 167904 (2001).

\bibitem{Shchukin} E. Shchukin, and W. Vogel, \prl \textbf{95}, 230502 (2005).

\bibitem{Biswas} G.S. Agarwal, and A. Biswas, New. J. Phys. \textbf{7}, 211 (2005).

\bibitem{Zubairy} M. Hillery, and M.S. Zubairy, \prl \textbf{96}, 050503 (2006).

\bibitem{Nha} H. Nha, S.-Y. Lee, S.-W. Ji, and M.S. Kim, \prl \textbf{108}, 030503 (2012).

\bibitem{Nha1} H. Nha, and H.J. Carmichael, \prl \textbf{93}, 020401 (2004).

\bibitem{Cerf} R. Garc\'ia-Patr\'on, J. Fiur\'a\u sek, N. J. Cerf, J. Wenger, R. Tualle-Brouri, and P. Grangier, \prl \textbf{93}, 130409 (2004).

\bibitem{PLLN12} J. Park, S.-Y. Lee, H.-W. Lee, and H. Nha, J. Opt. Soc. Am. B \textbf{29}, 906 (2012).

\bibitem{Welsch} T. Opatrny, G. Kurizki, and D. G. Welsch, \pra \textbf{61}, 032302 (2000).

\bibitem{Cochrane} P. T. Cochrane, T. C. Ralph, and G. J. Milburn, \pra \textbf{65}, 062306 (2002).

\bibitem{Olivares} S. Olivares, M. G. A. Paris, and R. Bonifacio, \pra \textbf{67}, 032314 (2003).

\bibitem{Kitagawa} A. Kitagawa, M. Takeoka, M. Sasaki, and A. Chefles, \pra \textbf{73}, 042310 (2006).

\bibitem{Dell} F. Dell'Anno, S.De Siena, L. Albano, and F. Illuminati, \pra \textbf{76}, 022301 (2007); 
F. Dell'Anno, S. De Siena, and F. Illuminati, \pra \textbf{81}, 012333 (2010).

\bibitem{Yang1} Y. Yang, and F.L. Li, \pra \textbf{80}, 022315 (2009).

\bibitem{Takahashi} H. Takahashi, J.S. Neergaard-Nielsen, M. Takeuchi, M. Takeoka, 
K. Hayasaka, A. Furusawa, and M. Sasaki, Nature Photonics \textbf{4}, 178 (2010).

\bibitem{Li} S.L. Zhang, and P. van Loock, \pra \textbf{82}, 062316 (2010).

\bibitem{Suyong} S.-Y. Lee, S.-W. Ji, H.-J. Kim, and H. Nha \pra \textbf{84}, 012302 (2011).

\bibitem{Carlos} C. Navarrete-Benlloch, R. Garc\'ia-Patr\'on, J.H. Shapiro, and N.J. Cerf, \pra \textbf{86}, 012328 (2012).

\bibitem{Agarwal} G.S. Agarwal, and K. Tara, \pra \textbf{43}, 492 (1991).

\bibitem{Zavatta} A. Zavatta, S. Viciani, and M. Bellini, Science \textbf{306}, 660 (2004).

\bibitem{Wenger} J. Wenger, R. Tualle-Brouri, and Ph. Grangier, \prl \textbf{92}, 153601 (2004).

\bibitem{Filip}  P. Marek and R. Filip, \pra \textbf{81}, 022302 (2010).

\bibitem{Usuga} M. A. Usuga, C. R. M\"uller, P. Marek, R. Filip, C. Marquardt, G. Leuchs, and U. L. Andersen, Nat. Phys. \textbf{6}, 767 (2010).

\bibitem{Jeffers} J. Jeffers, \pra \textbf{83}, 053818 (2011).

\bibitem{Hojoon}H.-J. Kim, S.-Y. Lee, S.-W. Ji, and H. Nha, \pra \textbf{85}, 013839 (2012).

\bibitem{Kim} M.S. Kim, J. Phys. B: At. Mol. Opt. Phys. \textbf{41}, 133001 (2008).

\bibitem{Yang} Y. Yang, and F.L. Li, J. Opt. Soc. Am. B \textbf{26}, 1363 (2009).

\bibitem{Lee} S.-Y. Lee, J. Park, S.-W. Ji, C.H.R. Ooi, and H.-W. Lee, J. Opt. Soc. Am. B \textbf{26}, 1532 (2009). 

\bibitem{Parigi} V. Parigi, A. Zavatta, M.S. Kim, and M. Bellini, Science \textbf{317}, 1890 (2007).

\bibitem{ZFB11} A. Zavatta, J. Fiur\'a\u sek, and M. Bellini, Nature Photon. \textbf{5}, 52 (2011).

\bibitem{Pawel} P. Kurzy\'nski, R. Ramanathan, A. Soeda, T.K. Chuan, and D. Kaszlikowski, New. J. Phys. \textbf{14}, 093047 (2012).

\bibitem{LNK13} S.-Y. Lee, C. Noh, and D. Kaszlikowski, arXiv: 1212.5338








\bibitem{Kim2} M.S. Kim, H. Jeong, A. Zavatta, V. Parigi, and M. Bellini, \prl \textbf{101}, 260401 (2008).

\bibitem{Zavatta1} A. Zavatta, V. Parigi, M.S. Kim, H. Jeong, and M. Bellini, \prl \textbf{103}, 140406 (2009). 

\bibitem{Park} J. Park, S.-Y. Lee, H.-J. Kim, and H.-W. Lee, New J. Phys. \textbf{12}, 033019 (2010);
H.-J. Kim, J. Park, and H.-W. Lee, J. Opt. Soc. Am. B \textbf{27}, 464 (2010).

\bibitem{SuYong} S.-Y. Lee, and H. Nha, \pra \textbf{82}, 053812 (2010).

\bibitem{Yong} S.-Y. Lee, and H. Nha, \pra \textbf{85}, 043816 (2012).

\bibitem{Changwoo} C.-W. Lee, J. Lee, H. Nha, and H. Jeong, \pra \textbf{85}, 063815 (2012). 

\bibitem{SJHH} S.-Y. Lee, J. Park, H.-W. Lee, and H. Nha, Opt. Express \textbf{20}, 14221 (2012).


\bibitem{Nobert}  N.~L{\"u}tkenhaus and S. M. Barnett, \pra \textbf{51}, 3340 (1995).

\bibitem{Braunstein} S.L. Braunstein, and H.J. Kimble, \prl \textbf{80}, 869 (1998).




\bibitem{Yurke} B. Yurke, and D. Stoler, \prl \textbf{57}, 13 (1986).

\bibitem{Mecozzi} A. Mecozzi, and P. Tombesi, \prl \textbf{58}, 1055 (1987).

\bibitem{Marek} P. Marek, H. Jeong, and M.S. Kim, \pra \textbf{78}, 063811 (2008).

\bibitem{Polzik} J.S. Neergaard-Nielsen, B.M. Nielsen, C. Hettich, K. Molmer, and E.S. Polzik, \prl \textbf{97}, 083604 (2006).

\bibitem{Grangier1} A. Ourjoumtsev, R. Tualle-Brouri, J. Laurat, and Ph. Grangier, Science \textbf{312}, 83 (2006).

\bibitem{Grangier2} A. Ourjoumtsev, H. Jeong, R. Tualle-Brouri, and Ph. Grangier, Nature(London) \textbf{448}, 784 (2007).

\bibitem{Sasaki1} H. Takahashi, K. Wakui, S. Suzuki, M. Takeoka, K. Hayasaka, A. Furusawa, and M. Sasaki, \prl \textbf{101}, 233605 (2008).

\bibitem{Sasaki2} M. Sasaki, M. Takeoka, and H. Takahashi, \pra \textbf{77}, 063840 (2008); M. Takeoka, H. Takahashi, and M. Sasaki, ibid. \textbf{77}, 062315(2008).

\bibitem{Grangier3} A. Ourjoumtsev, F. Ferreyrol, R. Tualle-Brouri, and Ph. Grangier, Nat. Phys. \textbf{5}, 189 (2009).

\bibitem{Gerrits} T. Gerrits, S. Glancy, T.S. Clement, B. Calkins, A.E. Lita, A.J. Miller, A.L. Migdall, S.W. Nam, R.P. Mirin, and E. Knill, \pra \textbf{82}, 031802 (2010).

\bibitem{BBPS} C.H. Bennett, H.J. Bernstein, S. Popescu, and B. Schumacher, \pra \textbf{53}, 2046 (1996).



\bibitem{Braunstein2} S.L. Braunstein, C.A. Fuchs, H.J. Kimble, and P. van Loock, \pra \textbf{64},
022321 (2001).

\bibitem{Marian} P. Marian, and T.A. Marian, \pra \textbf{74}, 042306 (2006).



\end{thebibliography}
\end{document}